\documentstyle[12pt,psfig]{article}
\begin{document}
\newcommand{\cd}{\makebox[0.08cm]{$\cdot$}}

\centerline{\bf $(\Omega \Omega)_{0^+}$ dibaryon productions in central Au+Au
collisions}
\centerline{\bf at RHIC energy $\sqrt {s_{NN}} =130$GeV}
\vskip 14pt
\centerline{X.-M. Xu}
\centerline{Nuclear Physics Division, Shanghai Institute of Nuclear Research}
\centerline{Chinese Academy of Sciences, P.O.Box 800204, Shanghai 201800,
            China}
\vskip 14pt
\centerline{P. Wang, Y. W. Yu}
\centerline{Institute of High Energy Physics,}
\centerline{Chinese Academy of Sciences,
Beijing  100039, China}

\begin{abstract}
\baselineskip=14pt
Based on the measured transverse mass spectra of $\pi^-$, $K^-$ and $\bar p$
at the RHIC energy $\sqrt {s_{NN}} =130$GeV, di-omega productions from
baryon-baryon reactions in hadronic matter are studied. Results about the
$(\Omega \Omega)_{0^+}$ total number per event
 show that  the deeply bound state
$(\Omega \Omega)_{0^+}$ can be observed at RHIC energies.
\end{abstract}
\leftline{PACS codes: 14.20.Pt,25.75.-q,25.75.Dw,}
\leftline{Keywords: dibaryon, hadronic matter}

\newpage
\leftline{\bf 1. Introduction}
\vspace{0.5cm}

Data measured at RHIC experiments have revealed intriguing new features
in ultrarelativistic nuclear collisions [1]. Various theoretical models have
undergone challenges from the experimental data and turn out to be adjusted.
The unexpected large values
of the elliptic flow signal $v_2$ can be accounted for by hydrodynamic
calculations [2] indicating perhaps complete early-time thermalization in
central Au+Au collisions at $\sqrt {s_{NN}}= 130$GeV per nucleon pair.
This stresses the collective  flow strongly affects hadrons in the
hadronic matter produced at the center-of-mass energy. Therefore,
any further predictions relying on the hadronic matter must involve available
hydrodynamic behaviors.

One of the interesting topics in hadronic physics is the search for dibaryon
bound states.
Since dibaryon is a kind of new matter, it can
provide a good place to examine the short-range quark-gluon
behaviour of QCD theory and also would open a new area for
studying many new physical phenomena we have not known before.
Even though the $H$ dibaryon predicted by Jaffe [3] has not been confirmed in 
hadron-hadron
collison experiments, recently, the search is extended to Au+Au collisions in
the AGS Experiment 896 [4]. Dibaryons involving $\Delta$ and nucleon were
 predicted in an adiabatic method [5].
Productions of short-lived dibaryons with strangeness, for instance,
$\Sigma^+ p$, $\Xi^0 p$, $\Xi^0 \Lambda$, $\Xi^0 \Xi^-$,   
have been  calculated  for Au+Au collisions at $\sqrt {s_{NN}}
=200$GeV and  may be identified as peaks in the invariant mass
spectra of final two baryons yielded from dibaryon decays [6].
Our present work is based on
a dynamical investigation on dibaryon structure made in
the framework of chiral $SU(3)$ model [7-9], in which the
coupling between  chiral fields and quarks is considered to
describe the non-perturbative QCD effect, and therefore
the huge data of  nucleon-nucleon scattering phase
shifts and  hyperon-nucleon cross sections can be reproduced
correctly. This investigation provided the di-omega
$(\Omega \Omega)_{0^+}$ as the most interesting structure [8].
It is deeply bound with the binding energy about 100MeV and its
mean life time is in order of $10^{-10}$ second because it can only
undergo through  weak decay. This charged dibaryon
could be easily
identified in experiments carrying  on ultrarelativistic nucleus-nucleus
collsions[8].

In order to find the deeply bound state $(\Omega
\Omega)_{0^+}$ in RHIC experiments, we calculate its yield
per event in central Au+Au  collisions at $\sqrt {s_{NN}} =130$GeV.
As we will see in the following, the
hydrodynamic equation with transverse flow is used and measured
data constitute inputs in all calculations to make results as reliable as
possible.
Cross sections for $(\Omega \Omega)_{0^+}$
yields in baryon-baryon scatterings are taken into account.

\vspace{0.5cm}
\leftline{\bf 2. Hadron distributions}
\vspace{0.5cm}
The hadron spectra measured in RHIC experiments allow us to analyse hadron
distributions evolving with time via the Cooper-Frye formula [10], hydrodynamic
equation and baryon number conservation. Assume hadron distributions are in the
Boltzmann form
\begin{equation}
f(\lambda,p,T)=g\lambda(\tau, \eta, r) e^{-p \cdot u /T}
\end{equation}
where $g$, $\lambda$, $p$ and $T$ are the degeneracy factor, fugacity,
four-momentum and temperature, respectively. The $u$ is the four-velocity of
fluid flow  given by the cylindrical polar coordinates $r$, $\phi$ and $z$
[11,12]. The fugacity $\lambda$ depends on the proper time $\tau$, space-time
rapidity $\eta$ and $r$. The momentum distribution of particle given by means
of the Cooper-Frye formula at freeze-out time $\tau_{fo}$ is
\begin{eqnarray}
\frac {d^2N(\tau_{fo})}{m_{\bot}dm_{\bot}dy} &  =  &
\frac {g\tau_{fo}}{(2\pi)^2} \int^{R(\tau_{fo})}_0 dr
\int^{\eta_{\rm max}}_{\eta_{\rm min}} d\eta
\int^{2\pi}_0 d\phi r \lambda (\tau, \eta, r)     \nonumber    \\
&  &
m_\bot \cosh ( y-\eta )
e^{-[m_\bot \cosh (y-\eta ) -p_\bot v_r \cos \phi ] /T\sqrt {1-v_r^2}}
\end{eqnarray}
where $N$, $y$, $p_\bot$, $m_\bot$ and $v_r$ are respectively the number,
rapidity, transverse momentum, transverse mass of hadron and
transverse velocity of fluid flow. The $R(\tau_{fo})$ is the transverse
radius of hadronic matter at freeze-out. The space-time rapidity spans from
a minimum $\eta_{\rm min}$ to a maximum $\eta_{\rm max}$.
Any measured spectra of hadrons allow one
to determine fugacities and transverse velocity at freeze-out. To gain
the velocity profile at any proper time $\tau$, the
hydrodynamic equation is used,
\begin{equation}
\partial_{\mu} T^{\mu \nu} =0
\end{equation}
where the energy-momentum tensor is
\begin{equation}
T^{\mu \nu} =(\epsilon + P) u^{\mu} u^{\nu} + P g^{\mu \nu}
\end{equation}
The $\epsilon$ and $P$ are energy density and pressure, respectively. Because
the dominant ingredient of hadronic matter is assumed to be pions, the equation
of state is $\epsilon =3P$. With the hydrodynamic equation for a
boost-invariant longitudinal  expansion along the $z$ axis and a cylindrically
symmetric transverse expansion [11], the transverse velocity $v_r (\tau,r)$ at
any time prior to freeze-out is obtained.

The dependence of hadron fugacity on time is obtained by the number
conservation
\begin{equation}
\partial_{\mu} (nu^\mu) =0
\end{equation}
which is followed by [13]
\begin{equation}
\partial_t \lambda +v_r \partial_r \lambda + \frac {\lambda}{\gamma T^3}
\partial_t (\gamma T^3) + \frac {\lambda v_r}{\gamma T^3}
\partial_r (\gamma T^3)
+\lambda \partial_r v_r +\lambda (\frac {v_r}{r} +\frac {1}{t}) =0
\end{equation}
where $\gamma =1/ \sqrt {1-v_r^2}$.

The equation for fugacity has the divergent term $\lambda v_r/r$ as $r \to 0$,
which causes difficulty
in solving  numerically the equation and  leads to nonphysical solutions. There
are four cases on curing this divergence:

\noindent
(1)$v_r \sim r^{\rm n}$ as $r \to 0$ with ${\rm n} \geq 1$;

\noindent
(2)$\lambda \sim r^{\rm m}$ as $r \to 0$ with ${\rm m} \geq 1$;

\noindent
(3)$v_r \sim r^{\rm n}$ and $\lambda \sim r^{\rm m}$ with ${\rm n} < 1$
and ${\rm m} < 1$ but ${\rm n}+{\rm m} < 1$ while $r \to 0$;

\noindent
(4)The term $\frac {\lambda v_r}{\gamma T^3} \partial_r (\gamma T^3)$ cancels
this term $\frac {\lambda v_r}{r}$ as $r \to 0$,
\begin{equation}
\frac {\lambda v_r}{\gamma T^3} \partial_r (\gamma T^3)
+\frac {\lambda v_r}{r}  =0
\end{equation}
which has the analytic solution
\begin{equation}
T=(\frac {C_1}{\gamma r})^{\frac {1}{3}}
\end{equation}
The constant $C_1$ is positive and in general depends on time. This
solution becomes unphysical if it is singular as $r \to 0$. To avoid this,
mandatorily force $\gamma r =C_2$ with $C_2$ being nonzero and positive. The
consequence is
\begin{equation}
v_r = \sqrt {1- (\frac {r}{C_2})^2}
\end{equation}
Then the condition $C_2 > r$ is required to give real transverse
velocity.

Before doing any calculations, we assume the hadronic matter freezing out at
$\tau_{fo} = 10$fm/c has a transverse radius $R(\tau_{fo})=10$fm. The
fugacity takes a form independent of $\eta$ and has the same dependence on $r$
for all sorts of hadrons except for different normalizations to fit rapidity
densities.
Given some trial functions for fugacities and velocities, the transverse mass
spectra
of $\pi^-$, $K^-$ and $\bar p$ at mid-rapidity in central Au+Au collisions at
$\sqrt {s_{NN}} =130$GeV in Ref. [14] are reproduced. We find
the following parametrizations showing $r$-dependences of fugacities and
transverse velocity profiles at freeze-out
\begin{equation}
\lambda_{fo} \propto r^4 e^{-0.7r},~~~~~~~~~~~~~~~~~~~~~~~~~~~
v_r=(\frac {9.65+r}{20.3})^3
\end{equation}
in the case (2) and
\begin{equation}
\lambda_{fo} \propto \frac {e^{0.2r}}{r^{0.3}},~~~~~~~~~~~~~~~~~~~~~~~~~~~
v_r=\sqrt {1-(\frac {r}{10.1})^2}
\end{equation}
in the case (4). Inserting the fugacities and transverse velocities into
Eq. (1), we obtain hadron distributions as $r \to 0$: $f \to 0$. Evolve
the hydrodynamic equation and the fugacity equation (6) back to a previous time
from
the freeze-out hypersurface, we still have $f \to 0$ as $r \to 0$ while the
transverse
velocity is kept as $r \to 0$ as  nonzero for the case (2) and the analytic
function in Eq. (9) remains with a $\tau$-dependence of $C_2$  for the case
(4). Such $f$ conflicts with the
finite amplitude of a real hadron distribution as $r \to 0$. The cases (2) and
(4) are therefore given up. Undoubtedly, nonzero hadron distributions beginning
with the formation
of hadronic matter may be found by seeking for a specific numerical method. But
we do not consider here this task since we have no intention of finding all
solutions of the hydrodynamic equation and the fugacity equation in coordinate
space and all hadron distributions via Eq. (2) on  the  feeze-out hypersurface.
Parametrizations for the fugacity and transverse velocity with the small $r$
behavior described in the case (3) are not found.

In the case (1) the fugacity and transverse velocity are not unique. The  first
set is
\begin{equation}
~~~~~~~~~~~~~~~~~~~~~  \lambda_{fo} \propto  \left \{
\begin{array}{ll}
1, & r < 7.5  \\
e^{-(r-7.5)^2}, &  r \geq 7.5   \\
\end{array}
\right.
~~~~~~~~~~~~~~~v_r = \frac {r}{10.4}~~~~~~~~~~~~~~~~~~
\end{equation}
The $v_r$ depends
linearly on $r$ like Ref. [15]. The second and third sets of parametrizations
are
\begin{equation}
~~~~~~~~~~~~~~~~~~~~~  \lambda_{fo} \propto  \left \{
\begin{array}{ll}
1,  & r < \pi  \\
e^{-(r-\pi )},  &  r \geq \pi   \\
\end{array}
\right.
~~~~~~~~~~~~~~~v_r = \tanh \frac {r}{4.8}~~~~~~~~~~~~~~~~~~
\end{equation}
\begin{equation}
~~~~~~~~~~~~~~~~~~~~~  \lambda_{fo} \propto  \left \{
\begin{array}{ll}
1,  & r < \frac {\pi}{2}  \\
e^{-(2r-\pi )},  &  r \geq \frac {\pi}{2}   \\
\end{array}
\right.
~~~~~~~~~~~~~~~v_r = \tanh \frac {r}{2.4}~~~~~~~~~~~~~~~~~~
\end{equation}
In fact, the third set is obtained from the second set by making the
replacement
$r \to 2r$. For $r > R(\tau_{fo})$ the fugacity in the exponential form gives
very
small value and has negligible contribution to the integration over $r$. Then
$\int^{R(\tau_{fo})}_0 dr = \int^{2R(\tau_{fo})}_0 dr$ is a very good
approximation and the two sets give the same dependence of
$d^2N/m_{\bot}dm_{\bot}dy$ on the transverse mass upon normalization. Actually,
the substitution $r
\to ar$ with $a >1$ leads to a class of fugacities and transverse velocities
\begin{equation}
~~~~~~~~~~~~~~~~~~~~~  \lambda_{fo} \propto  \left \{
\begin{array}{ll}
1,  & r < \frac {\pi}{a}  \\
e^{-(ar-\pi )},  &  r \geq \frac {\pi}{a}   \\
\end{array}
\right.
~~~~~~~~~~~~~~~v_r = \tanh \frac {ar}{4.8}~~~~~~~~~~~~~~~~~~
\end{equation}
which fits well the measured data of  $d^2N/m_{\bot}dm_{\bot}dy$ versus
$m_\bot$ for $\pi^-$, $K^-$
and $\bar p$. Since the three curves in Fig. 1 of Ref. [14] are plotted in
arbitrary units, the rapidity density $\frac {dN}{dy} = \int^{\infty}_m
m_{\bot} dm_\bot \frac {d^2 N}{m_{\bot} dm_\bot dy}$ is fitted to that
generated in HIJING/B$\rm \bar B$ [16] for $p$, $\bar p$, $n$ and $\bar n$,
respectively. In this way the magnitude of fugacity is completely fixed and
the minimum and maximum
of $\eta$ are individually equal to -4 and 4 for nucleon, -2.8 and 2.8 for
antinucleon. Meanwhile, a common freeze-out temperature $T=0.1$GeV is obtained
for $\pi^-$, $K^-$ and $\bar p$ as extracted in Ref. [14].

Define an average velocity
\begin{equation}
\bar {v}_r =\frac { \int^{R(\tau_{fo})}_0 dr v_r} {R(\tau_{fo})}
\end{equation}
The three sets of transverse velocity profiles in Eqs. (12)-(14) give $\bar
{v}_r \approx 0.48c, 0.67c, 0.83c$, respectively.

As already mentioned the measured data at RHIC energy might not well be
reproduced in some theoretical models. The produced $\Omega$ spectra is
unknown. Even the knowledge of
$\Omega^-$ spectra measured in CERN-SPS experiments is very limited. To
have a good sense on the $\Omega^-$ spectra, we utilize the experimental data
on the ratio of antiproton yield at RHIC to that at CERN-SPS. Since both the
antiproton and $\Omega^-$ at mid-rapidity can be created from junction and
pomeron [16,17],
we assume the ratio of $\Omega^-$ yield at RHIC to that at CERN-SPS is the same
as the ratio for antiproton. For rapidity densities of $\Omega^-$ and
antiproton,
\begin{equation}
\frac {dN_{\Omega^-, RHIC}}{dy} / \frac {dN_{\Omega^-, SPS}}{dy}
= \frac {dN_{\bar {p}, RHIC}}{dy} / \frac {dN_{\bar {p}, SPS}}{dy}
\end{equation}
It follows that
\begin{equation}
\frac {dN_{\Omega^-, RHIC}}{dy} = C \frac {dN_{\bar {p}, RHIC}}{dy}
\end{equation}
where
\begin{equation}
C= \frac {dN_{\Omega^-, SPS}}{dy} / \frac {dN_{\bar {p}, SPS}}{dy}
\end{equation}
is taken as a constant and determined at mid-rapidity  by
\begin{equation}
C=\int^{0.5}_{-0.5} dy \frac {dN_{\Omega^-, SPS}}{dy} / \int^{0.5}_{-0.5}
dy \frac {dN_{\bar {p}, SPS}}{dy}
\end{equation}
If the numerator is estimated from the values measured by WA97
Collaboration [18] and the denominator by NA49 Collaboration [19],  $C$ is
about equal to  0.01.

\vspace{0.5cm}
\leftline{\bf 3. Number of $(N\Omega)_{2^+}$}
\vspace{0.5cm}
The chiral $SU(3)$ quark model with the scalar nonet fields and pseudoscalar
nonet fields has been presented to study $(\Omega \Omega)_{0^+}$ productions
in baryon-baryon  scatterings [20,21].
To create such a deeply bound state, the electromagnetic process
\begin{equation}
\Omega +\Omega \to (\Omega \Omega )_{0^+} +\gamma
\end{equation}
and the mesonic process
\begin{equation}
\Omega +\Omega \to (\Omega \Omega )_{0^+} +\eta
\end{equation}
are considered first. Since
the state $(\Omega \Omega )_{0^+}$ is directly produced from two $\Omega$
baryons, the two reactions are called direct production processes.
In order to understand phyiscs well, the cross sections in Ref. [20] are
replotted as functions of the center-of-mass energy of the two incoming omegas,
$\sqrt s$, in Fig. 1.

The chiral $SU(3)$ quark model shows a weakly bound state $(N \Omega )_{2^+}$
composed of a nucleon and an omega. The $(N \Omega )_{2^+}$ dibaryon with
spin 2 and positive parity can be formed through the electromagnetic process
\begin{equation}
\Omega + N \to (N \Omega)_{2^+} + \gamma
\end{equation}
and the mesonic process
\begin{equation}
\Omega + N \to (N \Omega)_{2^+} + \pi
\end{equation}
This $(N \Omega )_{2^+}$ state enhances the yield of $(\Omega \Omega )_{0^+}$
by three-quark exchange processes during its collisions with $\Omega$ baryons
\begin{equation}
\Omega +(N \Omega)_{2^+} \to (\Omega \Omega )_{0^+} + N
\end{equation}
Such $(\Omega \Omega )_{0^+}$ dibaryon is yielded from the collision of
$\Omega$ with nucleon and the accompanying product $(N \Omega )_{2^+}$. The
reactions are called indirect production processes.
For present calculations it is better to redraw the cross sections in Ref. [21]
as functions of the center-of-mass energy of omega and nucleon in Fig. 2 and of
omega and $(N \Omega)_{2^+}$ in Fig. 3.

The number of $(N \Omega )_{2^+}$ contained in hadronic matter at a proper
time $\tau$  is
\begin{equation}
N_{(N\Omega )_{2^+}}= \int d^4 x
\int \frac {d^3 p_{\Omega}}{(2\pi)^3} f_{\Omega}(\lambda_{\Omega},p_{\Omega},T)
\int \frac {d^3 p_N}{(2\pi)^3} f_N(\lambda_N,p_N,T)
v_{rel} \sigma_{N\Omega} (\sqrt {s})
\end{equation}
which has the volume $\int
d^4 x = \int^{2\pi}_0 d\phi \int^{\tau}_{\tau_f} d\tau \tau
  \int^{R(\tau)}_0 dr r \int^{\eta_{\rm max}}_{\eta_{\rm min}} d\eta$ and where
$v_{rel}$ is the relative velocity of two scattering particles and
$\sigma_{N\Omega}$ is the sum of cross sections for the
reactions (23) and (24).
The number of dibaryon $(N \Omega)_{2^+}$ is calculated for a central Au+Au
collision at $\sqrt {s_{NN}}=130$GeV. Results are plotted in Fig. 4.
We take the hypothesis that $(N\Omega )_{2^+}$ thermalizes immediately in
hadronic matter once it is created.
Then the momentum distribution of $(N\Omega )_{2^+}$ satisfies the
Boltzmann form and the corresponding fugacity is  determined by
\begin{eqnarray}
N_{(N\Omega )_{2^+}}(\tau) &  =  &
\frac {g\tau}{(2\pi)^2} \int^{R(\tau)}_0 dr
\int^{\eta_{\rm max}}_{\eta_{\rm min}} d\eta
\int^{2\pi}_0 d\phi \int^{\infty}_{{\rm m}_{N\Omega}}  dm_{\bot}
\int^{+\infty}_{-\infty} dy
r \lambda_{N\Omega} (\tau, \eta, r)      \nonumber    \\ &  &
m_\bot^2 \cosh ( y-\eta )
e^{-[m_\bot \cosh (y-\eta ) -p_\bot v_r \cos \phi ] /T\sqrt {1-v_r^2}}
\end{eqnarray}
where ${\rm m}_{N\Omega}$ is the dibaryon mass. The dibaryon fugacity
$\lambda_{N\Omega}$  takes the same $r$-dependence as a baryon except for
normalization.

\vspace{0.5cm}
\leftline{\bf 4. Di-omega productions per event}
\vspace{0.5cm}
While the momentum distributions of $\Omega$ and $(N\Omega )_{2^+}$
are known, the di-omega yield per event is calculated via the
formula
\begin{equation}
N= \int d^4 x \int \frac {d^3 p_1}{(2\pi)^3} f_1(\lambda_1,p_1,T)
\int \frac {d^3 p_2}{(2\pi)^3} f_2(\lambda_2,p_2,T) v_{rel} \sigma_{\Omega
\Omega} (\sqrt {s})
\end{equation}
where $\int
d^4 x = \int^{2\pi}_0 d\phi \int^{\tau_{fo}}_{\tau_f} d\tau \tau$
 $\int^{R(\tau)}_0 dr r \int^{\eta_{\rm max}}_{\eta_{\rm min}} d\eta$
with $\tau_f=1.5$fm/c corresponding to the temperature $T \approx 0.19$GeV.
At
the moment of forming hadronic matter, the interacting glod-gold
system
has already expanded. Presumably, the $R(\tau)$ grows linearly with the proper
time from
$R(\tau_f)=7$fm  to $R(\tau_{fo})=10$fm. The $\sigma_{\Omega \Omega}$ include
cross sections for the reactions (21), (22) and (25).

Given a fugacity and transverse flow velocity, number of dibaryon $(\Omega
\Omega
)_{0^+}$ yielded in a central Au+Au collision at $\sqrt {s_{NN}}=130$GeV is
listed in Table 1. $N_1$, $N_2$ and $N_5$
are individually the number of $(\Omega \Omega )_{0^+}$ created through the
reactions $\Omega +\Omega \to (\Omega \Omega )_{0^+} +\gamma$,
$\Omega +\Omega \to (\Omega \Omega )_{0^+} +\eta$  and
$\Omega +(N \Omega)_{2^+} \to (\Omega \Omega )_{0^+} +N$. The total number
obtained in the work is
\begin{equation}
N_{tot}=N_1 +N_2 +N_5
\end{equation}
where $N_1$, $N_2$ and $N_5$ are calculated with the use of Eq. (28).

\vspace{0.5cm}
\leftline{\bf 5. Discussions and conclusions}
\vspace{0.5cm}
We have shown that different hadron distributions at freeze-out lead to
different $(\Omega \Omega)_{0^+}$ numbers produced in a Au+Au collision and
have
obtained that $\pi^-$, $K^-$ and $\bar p$ freeze out at the same temperature
$T=0.1$GeV while the radius of transverse extension of cylindrically symmetric
hadronic matter is 10fm. If the freeze-out temperature or transverse radius
changes, these distributions vary. The fugacities and transverse velocities in
Eqs. (14)  and (15)
result from the replacement $r \to 2r$ and $r \to ar$ in the parametrization
(13). However, the hydrodynamic equation and the fugacity equation (6) are not
invariant under
the replacement. This is one reason why the  number of $(\Omega \Omega)_{0^+}$
corresponding to the parametrization (13) is different from that to the other
parametrization (14). Another reason is that a factor $a^2$ accounts for the
variation of the integration $\int dr r$ under the replacement $r \to ar$.

In the indirect production processes the number of $(N \Omega)_{2^+}$ depends
on the freeze-out temperature, the final transverse radius and the final proper
time when
the hadronic matter becomes free particles.
The yield of $(\Omega \Omega)_{0^+}$ in the indirect production
processes is actually proportional to the square of the distributions of proton
and neutron. Fig. 4 exhibits very small number of $(N \Omega)_{2^+}$. However,
the cross sections for each of indirect production processes, especially the
exothermal reaction $\Omega +(N \Omega)_{2^+} \to (\Omega \Omega )_{0^+} +N$ in
Fig. 3,
are much larger than the cross sections for the direct production processes.
The amount of di-omega gained
in the indirect production processes eventually exceeds that in the direct
production
processes. Nevertheless, this is true only near freeze-out when the
$(N\Omega)_{2^+}$ number is relatively larger, and false near
hadronization  when the $(N\Omega)_{2^+}$ number is lack.

It is shown by the three curves in Fig. 4 that the $(N\Omega)_{2^+}$ dibaryon
number increases rapidly during the time interval from $\tau=1.5$fm/c to 3fm/c.
Almost all of $(N\Omega)_{2^+}$ are produced in the early time of hadronic
matter and sensitive to early-time thermalization. The variation of dibaryon
number with time is controlled by high
baryon densities in the early time of hadronic matter and later on low baryon
densities.
We are  also
interested in calculating the dependence of $(\Omega \Omega)_{0^+}$ number on
the proper time with, for instance,  the fugacity and transverse velocity in
Eq. (12). Unlike the $(N\Omega)_{2^+}$, we find that the di-omega number
changes slowly with time
 and almost all of $(\Omega \Omega)_{0^+}$ are produced in the
late time of hadronic matter. Even if the $(N\Omega)_{2^+}$ number increases
rapidly, its density rises up not fast because the matter volume increases.
Together
with the reducing density of $\Omega^-$ with time, the $(\Omega \Omega)_{0^+}$
is produced slowly in hadronic matter.

The reaction $\Omega +\Omega \to (\Omega \Omega)_{0^+} +\eta$ has the threshold
energy 3.776GeV higher than the value 3.344GeV for
$\Omega +\Omega \to (\Omega \Omega)_{0^+} +\gamma$. Peaks of cross sections are
shown in Fig. 1 for both reactions. Thereby the $\Omega$ has larger average
energy in the $\eta$ yield process than in the $\gamma$ yield process.
This considerably reduces
the $(\Omega \Omega)_{0^+}$ yield from the former reaction compared to the
latter by the momentum distribution of $\Omega$. However, the ratio  $N_2/N_1
\approx 2.5$ in Table 1 reverse the case.
This comes out of the following factors: the reaction for $\eta$ production has
larger cross section and a wider peak than for $\gamma$; higher relative
velocity of two
$\Omega$s is demanded for $\eta$ production;  $d^3p_1 d^3p_2$ in Eq. (28) is
greatly enhanced by the larger average momentum of $\Omega$ needed for
$\eta$ yield. Therefore, more $(\Omega \Omega)_{0^+}$ dibaryons are produced
in the $\eta$ yield process than in the $\gamma$ yield process.

We have not estimated di-omega yields in the expansion process from the moment
of initial nucleus-nucleus collision to the time $\tau_f =1.5$fm/c when
hadronic matter appears.
During this period the central Au+Au collision at $\sqrt {s_{NN}}=130$GeV
may lead to deconfined gluons and quarks  in
nonequilibrium, if we believe at CERN-SPS energy the Pb+Pb collisions are on
the boundary of phase transition involving hadronic and partonic degrees of
freedom. While the quarks and gluons hadronize
at $\tau_f$, it is possible for six strange quarks to combine into a dibaryon
$(\Omega \Omega)_{0^+}$ as well as six light quarks to form $(N
\Omega)_{2^+}$. Then the number of $(N \Omega)_{2^+}$ at $\tau_f =1.5$fm/c is
not zero and moves upwards these curves in Fig. 4.

It is unavoidable to address in-medium effect on the production of $(\Omega
\Omega)_{0^+}$. The mass of strange quark is reduced and the quark-quark
potential is screened in hadronic matter at finite temperature. Obviously the
screening decreases the binding energy of $(\Omega \Omega)_{0^+}$. However, the
kinetic-energy contribution increases the binding energy by the dropping mass
of strange quark. We can still expect an appreciable binding energy for
$(\Omega \Omega)_{0^+}$ in hadronic matter with reference to the large binding
energy of free $(\Omega \Omega)_{0^+}$. In addition, hadronic matter modifies
the relative-motion  wave function of two omegas inside the dibaryon and the
coupling of $\eta$ to $\Omega$ in $\Omega +\Omega \to (\Omega \Omega)_{0^+} +
\eta$ and the coupling of $\pi$ to $N$ in $\Omega +N \to (N\Omega)_{2^+} +\pi$.
These result in different cross sections for the direct and indirect production
processes in comparison to Figs. 1-3 and different $(\Omega \Omega)_{0^+}$
number yielded relative to Table 1. Exploring in-medium effect remains in a
future work.

In conclusion, we have obtained the  number of
dibaryon $(\Omega \Omega)_{0^+}$ produced in a central Au+Au collision at
$\sqrt
{s_{NN}} =130$GeV is of the order of $10^{-9} \sim 10^{-7}$. Even if the number
is very small,
enough events will be accumulated to observe
the yield in the coming years of running of RHIC machine. Besides, we have
shown: the mesonic processes have larger contributions than the electromagnetic
processes; at feeze-out the indirect production processes have larger
contributions than the direct production processes; di-omegas
are mostly produced
in late-time hadronic matter.

\vspace{0.5cm}
\leftline{\bf Acknowledgements}
\vspace{0.5cm}
We wish to thank X. N. Wang, N. Xu and Z. Y. Zhang for useful helps.
This work was supported
by the National Natural Science Foundation of China,
 the fund of Science and Technology Committee of Shanghai and the CAS 
Knowledge Innovation Project No. KJCX2-SW-N02.

\newpage
\centerline{\bf References}
\vskip 14pt
\leftline{[1]For reviews see {\it Proceedings of Quark Matter 2001}, SUNY,
Jan. 15-20. }
\leftline{[2]STAR Collaboration, K. H. Ackermann {\it et al.},  Phys. Rev.
Lett. 86(2001)402;}
\leftline{~~~P. F. Kolb, U. Heinz, P. Huovinen, K. J. Eskola, K. Tuominen,
Nucl. Phys.}
\leftline{~~~A696(2001)197;}
\leftline{~~~D. Teaney, J. Lauret, E. V. Shuryak, Nucl. Phys. A698(2002)479.}
\leftline{[3]R. L. Jaffe, Phys. Rev. Lett. 38(1977)195.}
\leftline{[4]E896 Collaboration, H. Caines {\it et al.}, Nucl. Phys.
A661(1999)170c.}
\leftline{[5]F. Wang, J. Ping, G. Wu, L. Teng, T. Goldman, Phys. Rev. 
C51(1995)3411;}
\leftline{~~~T. Goldman, K. Maltman, G. J. Stephenson Jr., J. Ping, F. Wang,
Mod. Phys. Lett.}
\leftline{~~~A13(1997)59.}
\leftline{[6]J. Schaffner-Bielich, R. Mattiello, H. Sorge, Phys. Rev. Lett.
84(2000)4305.}
\leftline{~~~S. D. Paganis, G. W. Hoffmann, R. L. Ray, J.-L. Tang, T. Udagawa,
R. S. Longacre,}
\leftline{~~~Phys. Rev. C62(2000)024906.}
\leftline{[7]Z. Y. Zhang, Y. W. Yu, L. R. Dai, High Energy Phys.  Nucl.
Phys. 20(1996)363;}
\leftline{~~~Z. Y. Zhang, Y. W. Yu, P. N. Shen, L. R. Dai, A. Faessler and  U.
Straub, Nucl. Phys.}
\leftline{~~~A625(1997)59.}
\leftline{[8]Y. W. Yu, Z. Y. Zhang, X. Q. Yuan, Commun. Theor.
Phys. 31(1999)1;}
\leftline{~~~Z. Y. Zhang, Y. W. Yu, C. R. Ching, T. H. Ho, Z. D. Lu, Phys. Rev.
C61(2000)065204.}
\leftline{[9]Q. B. Li, P. N. Shen, Z. Y. Zhang, Y. W. Yu, Nucl. Phys.
A683(2001)487.}
\leftline{[10]F. Cooper, G. Frye, Phys. Rev. D10(1974)186.}
\leftline{[11]H. von Gersdorff, L. McLerran, M. Kataja, P. V. Ruuskanen, Phys.
Rev. D34(1986)794.}
\leftline{[12]E. Schnedermann, J. Sollfrank, U. Heinz, Phys. Rev.
C48(1993)2462.}
\leftline{[13]D. K. Srivastava, M. G. Mustafa, B. M$\rm \ddot u$ller, Phys.
Rev. C56(1997)1064.}
\leftline{[14]N. Xu, M. Kaneta, in {\it Proceedings of Quark Matter 2001},
SUNY, Jan. 15-20.}
\leftline{[15]T. S. Bir$\rm {\acute o}$, Phys. Lett. B487(2000)133.}
\leftline{[16]S. E. Vance, M. Gyulassy, X.-N. Wang, Phys. Lett. B443(1998)45;}
\leftline{~~~~~S. E. Vance, M. Gyulassy, Phys. Rev. Lett. 83(1999)1735.}
\leftline{[17]D. Kharzeev, Phys. Lett. B378(1996)238.}
\leftline{[18]WA97 Collaboration, E. Andersen {\it et al.}, Phys. Lett.
B449(1999)401;}
\leftline{~~~~~J. Phys. G25(1999)171.}
\leftline{[19]NA49 Collaboration, F. Sikl$\rm {\acute e}$r {\it et al.}, Nucl.
Phys. A661(1999)45c.}
\leftline{[20]Y. W. Yu, P. Wang, Z. Y. Zhang, C. R. Ching, T. S. Ho, Commun.
Theor. Phys. }
\leftline{~~~~~35(2001)553.}
\leftline{[21]Y. W. Yu, P. Wang, Z. Y. Zhang, C. R. Ching, T. S. Ho and L. Y. 
Chu,}
\leftline{~~~~~High Energy Phys.  Nucl. Phys. (to be published).}

\newpage
\centerline {\bf Figure captions}
\vskip 14pt
\noindent
Fig. 1. The left  and right panels show cross sections for $\Omega +\Omega \to
(\Omega \Omega)_{0^+} +\gamma$ and $\Omega +\Omega \to
(\Omega \Omega)_{0^+} +\eta$, respectively,
as functions of the center-of-mass energy of two incident omegas.
\vspace{0.2cm}

\noindent
Fig. 2.
Dotted, dot-dashed and solid lines in the left panel show cross sections for
$\Omega +p  \to (p \Omega)_{2^+} +\gamma$,
 $\Omega + n \to (n \Omega)_{2^+} +\gamma$ and their sum,
respectively. The right panel shows the cross section for $\Omega + N \to
(N \Omega)_{2^+} +\pi$ where $N$ includes proton and neutron.
\vspace{0.2cm}

\noindent
Fig. 3.
Cross section for $\Omega + (N \Omega)_{2^+} \to (\Omega \Omega)_{0^+} +N$
as a function of the center-of-mass energy $\sqrt s$. The $N$ includes proton
and neutron.
\vspace{0.2cm}

\noindent
Fig. 4.
Solid, dashed and dot-dashed lines are the
numbers of $(N\Omega)_{2^+}$ produced with the parametrizations for fugacities
and transverse velocities in Eqs. (12)-(14), respectively.
\vspace{0.2cm}

\newpage
\vskip 14pt
\begin{table}
\caption{Number of $(\Omega \Omega )_{0^+}$ produced per event with the three
sets of fugacities and transverse velocities in  Eqs. (12)-(14).}
\begin{center}
\begin{tabular}{c|c|c|c}
\hline\hline
{\sl }      & {\sl ${\rm Eq.} (12)$}
& {\sl ${\rm Eq.} (13)$  }
& {\sl ${\rm Eq.} (14)$  }    \\
\hline
$N_1$ & {$9.29 \times 10^{-10}$} & {$2.82 \times 10^{-9}$} & {$1.13 \times
10^{-8}$}    \\
$N_2$ & {$2.27 \times 10^{-9}$} & {$7.57 \times 10^{-9}$} & {$2.92 \times
10^{-8}$}    \\
$N_5$ & {$3.43 \times 10^{-9}$} & {$3.6 \times 10^{-8}$} & {$5.95 \times
10^{-7}$}    \\
$N_{tot}$ & {$6.63 \times 10^{-9}$} & {$4.64 \times 10^{-8}$} & {$6.35 \times
10^{-7}$}    \\
\hline\hline
\end{tabular}
\end{center}
\end{table}

\newpage
\begin{figure}[t]
  \begin{center}
    \leavevmode
    \parbox{\textwidth}
           {\psfig{file=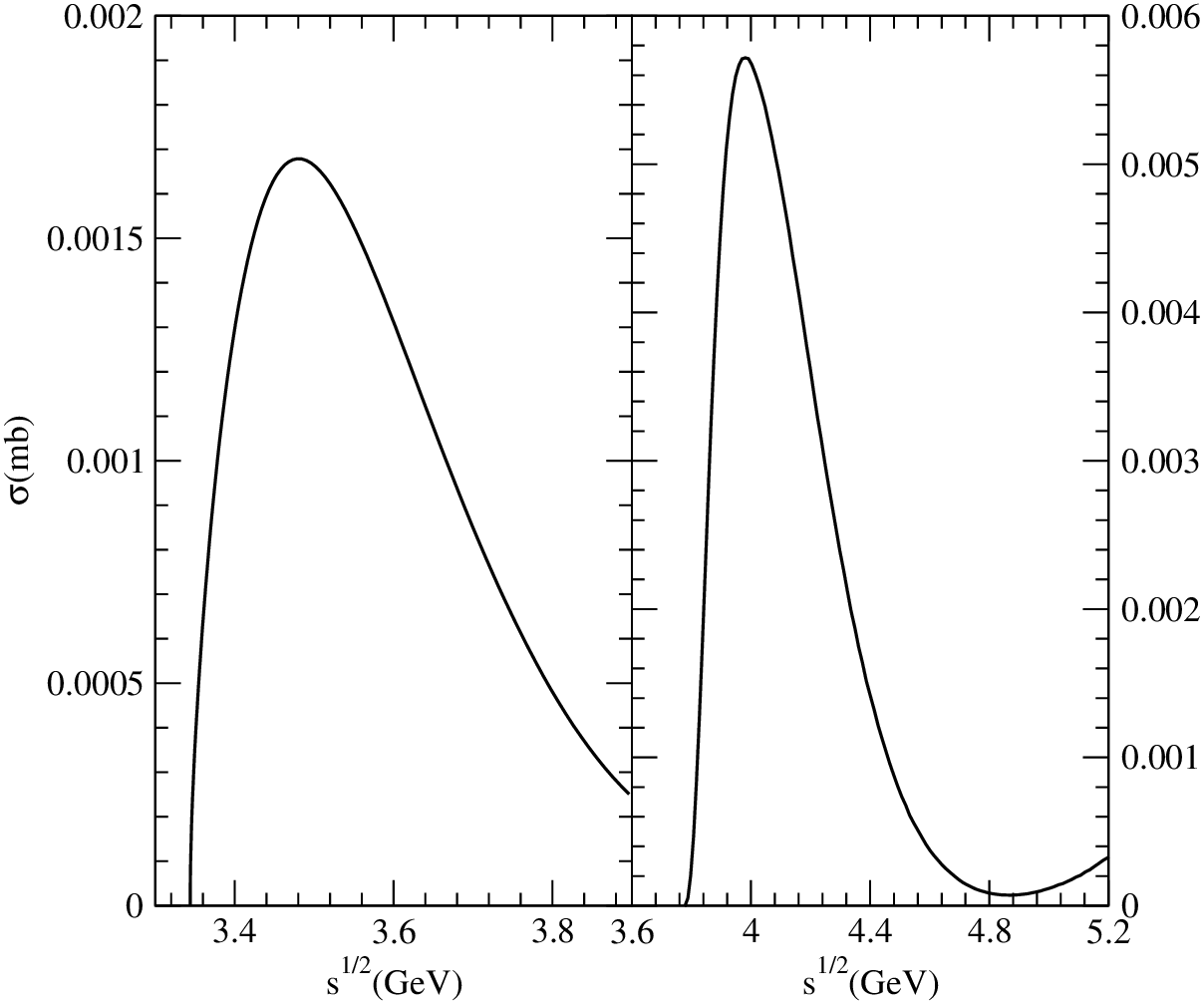,width=\textwidth,angle=0}}
  \end{center}
\caption{}
\label{fig1}
\end{figure}

\newpage
\begin{figure}[t]
  \begin{center}
    \leavevmode
    \parbox{\textwidth}
           {\psfig{file=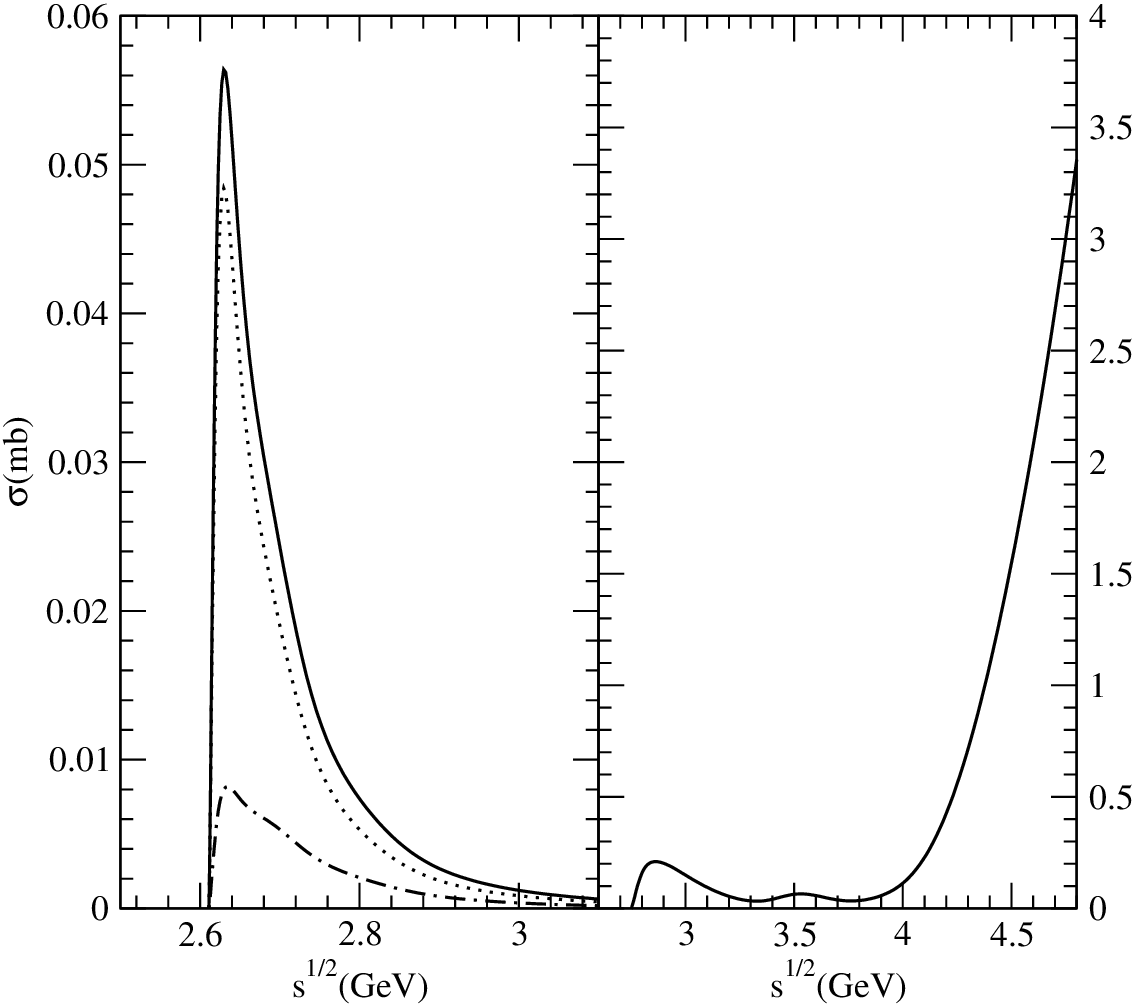,width=\textwidth,angle=0}}
  \end{center}
\caption{}
\label{fig2}
\end{figure}

\newpage
\begin{figure}[t]
  \begin{center}
    \leavevmode
    \parbox{\textwidth}
           {\psfig{file=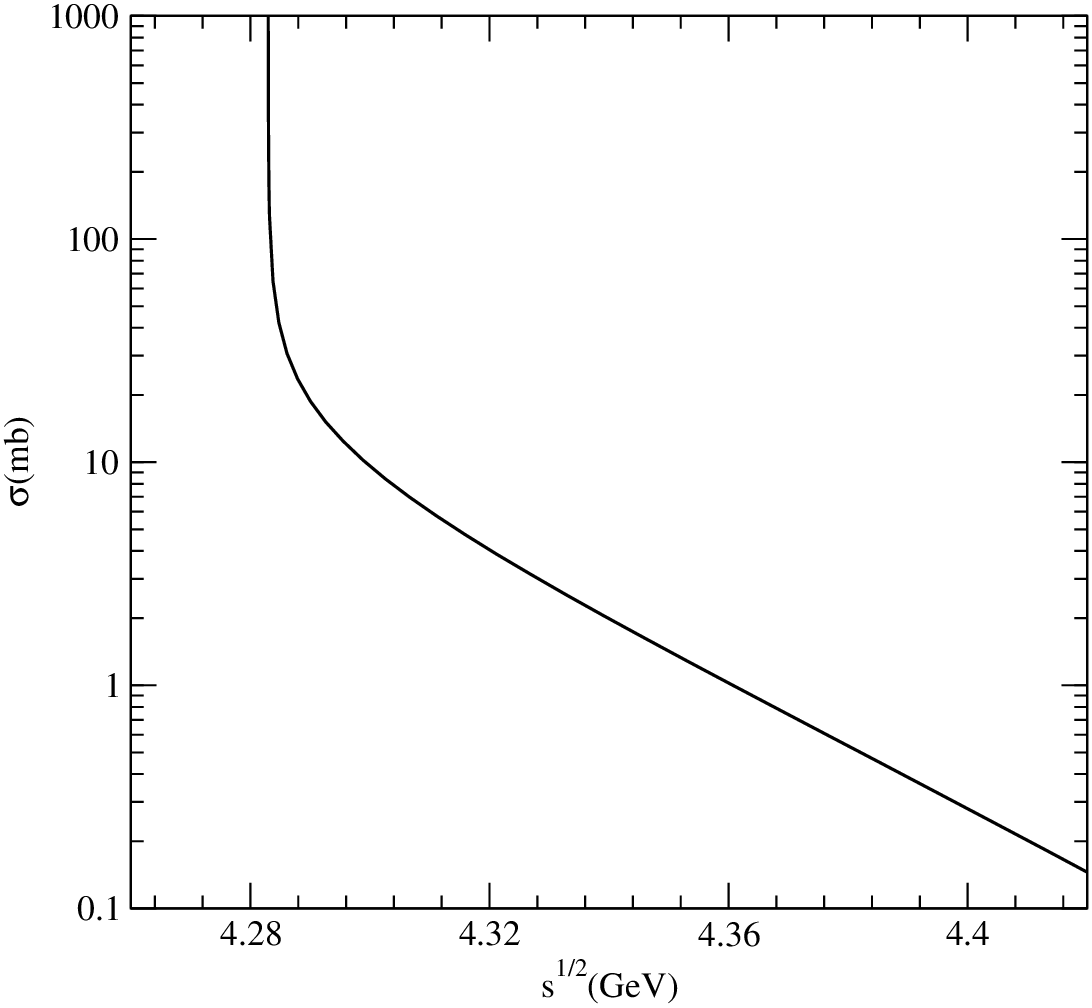,width=\textwidth,angle=0}}
  \end{center}
\caption{}
\label{fig3}
\end{figure}

\newpage
\begin{figure}[t]
  \begin{center}
    \leavevmode
    \parbox{\textwidth}
           {\psfig{file=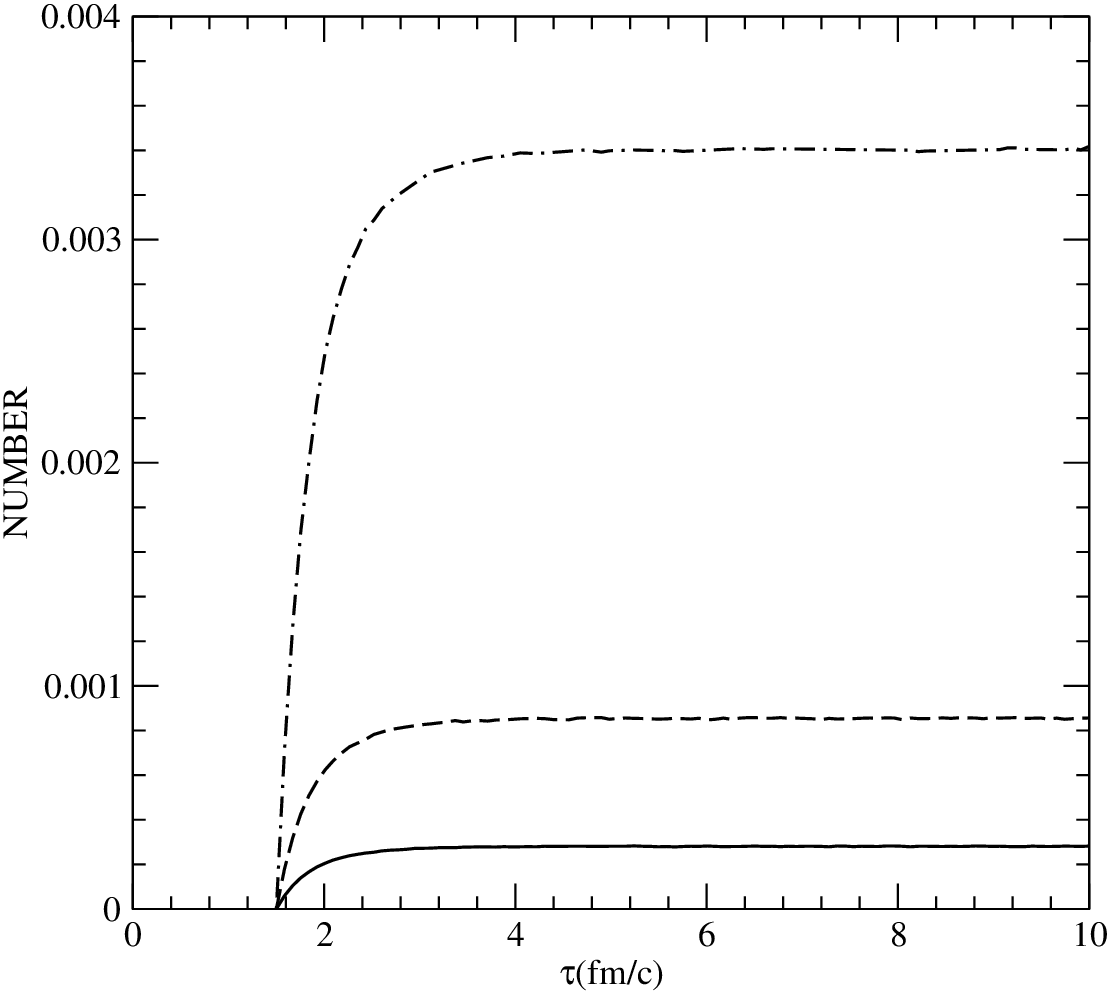,width=\textwidth,angle=0}}
  \end{center}
\caption{}
\label{fig4}
\end{figure}

\end{document}